\documentclass{article}

\usepackage{url} % for URLs in titles/notes

\usepackage[numbers,sort&compress]{natbib}

\usepackage{amsmath}
\usepackage{float}

\usepackage{arxiv}

\usepackage[utf8]{inputenc} % allow utf-8 input
\usepackage[T1]{fontenc}    % use 8-bit T1 fonts
\usepackage{hyperref}       % hyperlinks
\usepackage{url}            % simple URL typesetting
\usepackage{booktabs}       % professional-quality tables
\usepackage{amsfonts}       % blackboard math symbols
\usepackage{nicefrac}       % compact symbols for 1/2, etc.
\usepackage{microtype}      % microtypography
\usepackage{lipsum}
\usepackage{graphicx}
\graphicspath{ {./images/} }

\title{Cloud Investigation Automation Framework (CIAF): \\ An AI-Driven Approach to Cloud Forensics} %%%%%%%%%%%%
\author{
Dalal Alharthi\thanks{University of Arizona, \texttt{dalharthi@arizona.edu}} 
\\
Ivan Roberto Kawaminami Garcia\thanks{University of Arizona, \texttt{kawaminami@arizona.edu}}
}

\makeatletter
\@ifundefined{isaccepted}{\def\isaccepted{}}{}% trick the style into "accepted" mode
\makeatother
\ClearShipoutPicture % remove anything already queued by eso-pic

\begin{document}

\maketitle

\let\thefootnote\relax
\footnotetext{This is a preprint of a paper accepted to the International Conference on Digital Forensics and Cyber Crime (ICDF2C 2025).}

\begin{abstract}
Large Language Models (LLMs) have gained prominence in various domains, including cloud security and forensics. However, cloud forensic investigations still heavily rely on manual analysis, making them time-consuming and error-prone. LLMs can mimic human reasoning, providing a pathway to automating and enhancing cloud forensics log analysis. To address these challenges, we introduce the Cloud Investigation Automation Framework (CIAF), an ontology-driven framework that systematically investigates cloud forensic logs while improving analysis efficiency and accuracy. The framework standardizes user inputs through semantic validation, eliminating ambiguity and ensuring consistency in log interpretation. This validation mechanism not only enhances the quality of data analysis but also ensures that forensic investigators can rely on accurate, standardized information for making decisions. To evaluate its security and performance, we conducted an experiment analyzing Microsoft Azure logs containing distinct ransomware-related events. By simulating these attacks and assessing the framework’s impact, our results demonstrate a significant improvement in ransomware detection, achieving precision, recall, and F1 scores of approximately 93\%. This performance suggests that the CIAF is capable of significantly enhancing the detection of cyber threats in cloud environments. Furthermore, the framework’s modular and adaptable design ensures its applicability beyond ransomware, making it a robust solution for investigating a wide range of cyberattacks. By laying the foundation for standardized forensic methodologies and informing future developments in AI-driven forensic automation, this work highlights the pivotal role of deterministic prompt engineering and ontology-based validation in enhancing cloud forensic investigations. These advancements not only improve the effectiveness of cloud security but also pave the way for more efficient, automated forensic workflows in the future. 
\end{abstract} %%%%%%%%%

\def\UrlBreaks{\do\/\do-}

\def\BibTeX{{\rm B\kern-.05em{\sc i\kern-.025em b}\kern-.08em
    T\kern-.1667em\lower.7ex\hbox{E}\kern-.125emX}}

%in IoT or in hybrid cloud infrastructures


%\textit{name of the organization (of Aff.)}\\
%\and

\maketitle

\noindent\textbf{Keywords:} 
Cloud Forensics, AI-Driven Log Analysis, Ontology-Based Framework, 
Forensic Automation, Ransomware Detection, Large Language Models (LLMs), 
Azure Cybersecurity

\section{Introduction}
\label{intro}

Large Language Models (LLMs) have demonstrated remarkable advancements across diverse applications, including cloud security and digital forensics. Their ability to mimic human reasoning enables automation in threat detection and incident response \cite{derner2024taxonomy, chernyshev2024forensic}. However, cloud forensic investigations often remain manual and time-consuming, particularly in log analysis \cite{akula2024cloudtools, purnaye2022cloudforensics}. While research has primarily emphasized LLM scalability and performance, their potential role in enhancing cloud forensic investigations has not been largely explored.

Cloud environments remain vulnerable to ransomware attacks that exploit misconfigurations and weak security policies, disrupting operations and complicating forensic investigations through encryption and obfuscation techniques \cite{akula2024cloudtools, reshmi2021information}. Recent studies have systematically categorized cloud forensics challenges and analyzed adversarial attack patterns \cite{purnaye2022cloudforensics, mishra2012cloudforensics}. While automated forensic analysis tools exist \cite{zhu2023promptbench, schulhoff2023hackaprompt}, they remain reactive and lack structured ontology-driven validation. Emerging AI-driven frameworks, such as LangGraph \cite{langgraph}, AutoGen \cite{autogen}, and CrewAI \cite{crewAI}, offer promising advancements, but their forensic applications remain limited.

To address these challenges, we introduce the Cloud Investigation Automation Framework (CIAF), an ontology-driven system designed to standardize and validate cloud forensic logs, enhancing analysis efficiency and accuracy. Recent studies have emphasized the need for structured methodologies to improve forensic accuracy and automation \cite{al-mugerrn2023metamodeling, mishra2012cloudforensics}.  By framing forensic log analysis within causal reasoning and structured AI-driven validation, CIAF establishes a scalable approach to cloud forensic investigations.

The effectiveness of ontology-driven forensic analysis lies in its ability to impose structured constraints on cloud logs, reducing the noise and ambiguity often present in unstructured forensic datasets \cite{zawoad2013cloudforensics}. This approach aligns with robustness frameworks \cite{madry2018towards, carlini2017evaluating}, where structured constraints enhance forensic accuracy by filtering out irrelevant or misleading log events. Additionally, enforcing causal dependencies between attack patterns and forensic insights allows for systematic validation through causal inference frameworks \cite{pearl2009causality}. Understanding these interactions is crucial for quantifying forensic effectiveness and assessing generalization trade-offs in AI-driven investigations \cite{zhou2024algorithmic, li2023latent}.

Building on these foundations, we conducted experiments on Microsoft Azure cloud logs, containing distinct ransomware-related events. By simulating ransomware attacks and assessing the impact of deploying CIAF, our results demonstrate a significant improvement in forensic accuracy, achieving precision, recall, and F1 scores of approximately 93\%. These findings confirm CIAF’s ability to enhance cloud forensic investigations, automate log analysis, and improve ransomware detection. This proactive approach not only strengthens forensic capabilities but also establishes a foundation for scalable, modular solutions applicable to broader cybersecurity investigations.

The integration of AI-driven methodologies into cloud forensics presents critical research challenges and opportunities. How can ontology-driven LLM analysis optimize forensic accuracy without compromising system performance? To what extent can these methods scale to handle the complexities of dynamic cloud environments? Furthermore, how can structured AI-driven validation enhance the reliability of forensic investigations in high-stakes cybersecurity incidents? Addressing these questions will advance the field by bridging gaps in automation, accuracy, and scalability, ultimately contributing to more efficient and trustworthy cloud forensic frameworks.

The remainder of this paper is structured as follows: Section \ref{relatedWork} reviews related work on cloud forensics, ontology-based methodologies, LLM applications in cybersecurity, and AI-driven ransomware detection. Section \ref{CIAF} introduces the proposed Cloud Investigation Automation Framework (CIAF), detailing its architecture, ontology design, and forensic automation capabilities. Section \ref{AzureCaseStudy} presents the experimental setup, including data collection from Microsoft Azure logs, the evaluation methodology, and CIAF’s performance in forensic log analysis and ransomware detection. Section \ref{discussion} discusses key findings, limitations, and directions for future research. Finally, Section \ref{conc} concludes the paper.

%%%%%%%%%%%%%%%%%%%%%%%%%%%%%%

\section{Related Work and State of the Art}
\label{relatedWork}

Before the rise of AI-driven cloud forensics, research primarily focused on traditional forensic methodologies, which relied on manual log analysis and rule-based approaches to investigate cyber incidents. Foundational work by Goodfellow et al. \cite{goodfellow2015explaining} introduced adversarial examples, demonstrating how small perturbations in input data could cause deep learning models to misclassify. Building on this, Carlini and Wagner \cite{carlini2017towards} developed stronger attack methods and evaluated countermeasures, revealing persistent vulnerabilities in deep networks. In parallel, advances in adversarial robustness focused on certified defenses, such as randomized smoothing \cite{cohen2019certified}, which provides probabilistic guarantees of model resilience under adversarial perturbations. Privacy concerns also emerged as a critical research area, with Differential Privacy \cite{dwork2006calibrating} establishing formalized mechanisms to protect data while maintaining utility. These foundational studies set the stage for evolving research into the vulnerabilities of complex, high-dimensional ML systems. As scaling continues to drive AI performance, recent work suggests that structured learning approaches offer alternative pathways to enhancing security \cite{snell2024scaling}.

\subsection{LLMs in Cybersecurity and Digital Forensics}

With the rise of LLMs and Generative AI, new security risks have emerged, particularly prompt injection attacks, which manipulate the natural language flexibility of LLMs to produce unintended outputs. Recent work has systematically evaluated these attacks, highlighting their systemic risks in multi-agent settings \cite{liu2023prompt, liu2024automatic}. In multi-agent LLM environments, research has shown that manipulated prompts can propagate cascading failures, affecting autonomous decision-making in critical infrastructures such as transportation networks and cloud security systems \cite{ju2024flooding}. Existing mitigation strategies highlight the importance of structured defenses in distributed environments \cite{zhang2022security}. However, ensuring robust security in large-scale, collaborative AI deployments remains a significant challenge, requiring a deeper integration of theoretical guarantees for adversarial robustness and causality-aware security frameworks \cite{muliarevych2024security}.

Recent studies have demonstrated the growing role of LLMs in cybersecurity, particularly in automating security analysis and enhancing forensic readiness. Yao et al. \cite{yao2025llms} provided a systematic review of LLM applications in cybersecurity, highlighting their ability to process complex security logs, detect anomalies, and generate actionable forensic insights. As LLMs continue to be integrated into forensic workflows, ensuring the integrity and auditability of their outputs becomes a crucial area of research \cite{bendiab2024forensic, galhotra2024logging, alharthi2023secure}.

\begin{figure*}[t]
%\vskip 0.2in
\begin{center}
\centerline{\includegraphics[width=\textwidth]{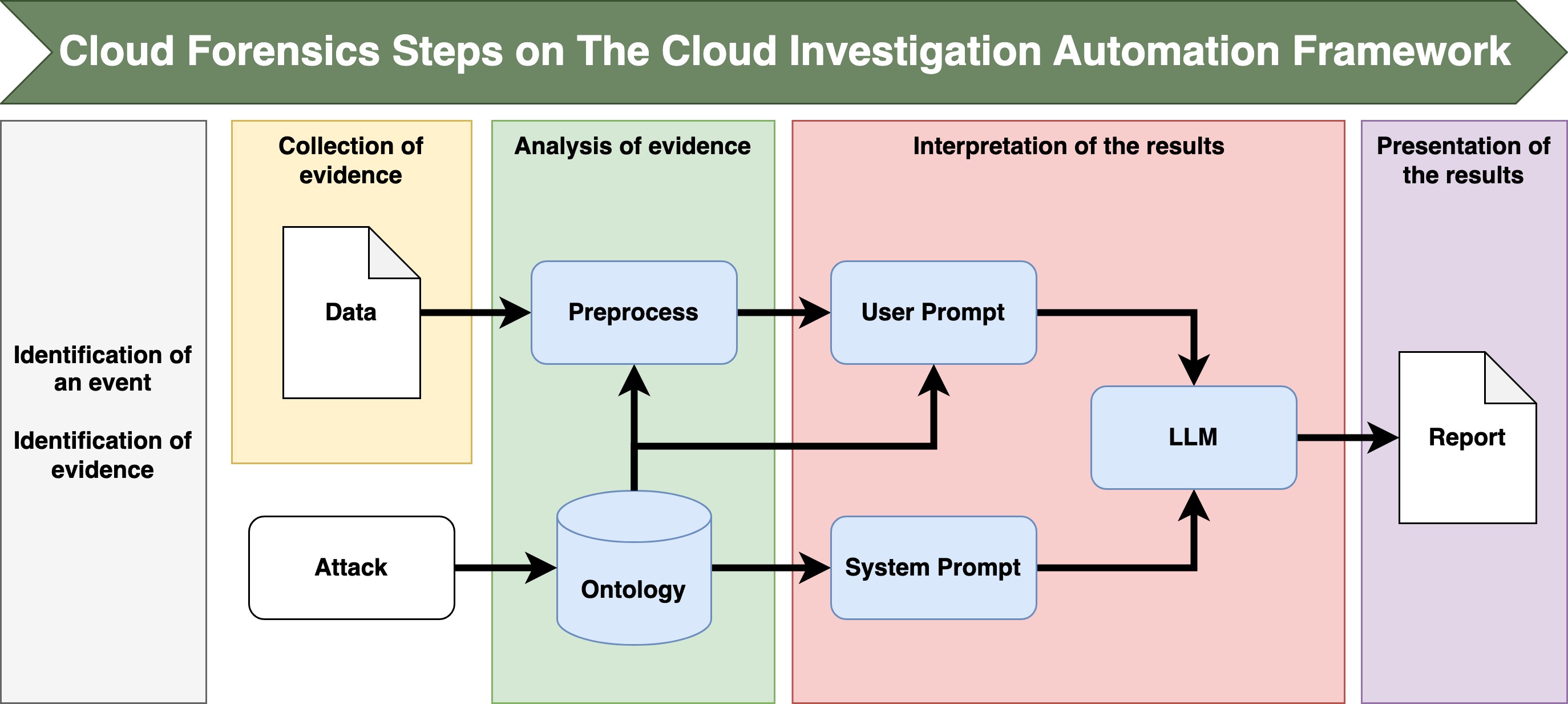}}
\caption{Flow diagram that shows how framework is implemented based on the cloud forensics steps.}
\label{Framework}
\end{center}
\vskip -0.2in
\end{figure*}

\subsection{Ontology-Driven Cloud Forensics}

Traditional cloud forensics has long faced challenges in processing large volumes of unstructured log data. Security Information and Event Management (SIEM) systems have been widely used for detecting suspicious activities by correlating log data from various sources \cite{denning1987intrusion, kent2006guide}. However, these rule-based methods often struggled to keep pace with evolving cyber threats and required significant manual intervention to update rules and models. Ontology-based frameworks have been proposed to address this challenge by providing structured representations of forensic data, improving accuracy and efficiency in forensic investigations. For example, Rouached et al. \cite{rouached2018ontology} introduced an ontology-driven approach for web services logs, enabling better cyber-attack detection. Similarly, the Cloud Forensic Readiness as a Service (CFRaaS) model \cite{li2023cloudforensic} emphasizes proactive accumulation of digital evidence, leveraging ontological structures to enhance forensic readiness in cloud environments.

Beyond forensic readiness, ontology engineering has become a critical tool in cyber threat intelligence. A recent study by Bratsas et al. \cite{bratsas2024knowledge} highlights how knowledge graphs and semantic web tools can improve forensic investigations by structuring cyber threat data. This approach reduces investigative ambiguity, enhances forensic accuracy, and ensures a standardized methodology for detecting and responding to cyber incidents \cite{alharthi2024cloud}.

\subsection{AI-Based Ransomware Detection and Cloud Security}

Despite advancements in forensic automation, cloud environments remain highly vulnerable to ransomware attacks, which exploit misconfigurations and weak security policies \cite{akula2024cloudtools}. Ransomware not only disrupts operations but also complicates forensic investigations by leveraging encryption and obfuscation techniques that hinder log analysis and evidence collection \cite{reshmi2021information}. Recent studies have systematically categorized cloud forensics challenges and analyzed adversarial attack patterns \cite{purnaye2022cloudforensics, mishra2012cloudforensics}. While automated forensic analysis tools exist \cite{zhu2023promptbench, schulhoff2023hackaprompt}, they remain reactive and lack structured ontology-driven validation. Emerging AI-driven frameworks, such as LangGraph \cite{langgraph}, AutoGen \cite{autogen}, and CrewAI \cite{crewAI}, offer promising advancements, but their forensic applications remain limited.

The application of AI-based ransomware detection is rapidly evolving. Ahmad et al. \cite{ahmad2025ai} conducted a comprehensive review of AI-driven techniques for identifying ransomware in cloud environments, emphasizing the need for anomaly detection and predictive modeling. In addition, machine learning (ML) approaches have been explored for forensic automation. Alhawi et al. \cite{alhawi2025trusted} proposed a meta-feature-based detection method leveraging volatile memory analysis to improve ransomware identification in private cloud infrastructures. These studies demonstrate how AI-powered solutions can proactively detect and mitigate ransomware attacks before they cause significant damage.

\subsection{LLM-Based Cloud Forensic Investigation Tools}

The integration of LLMs into digital forensics is an emerging field, with studies showing their potential to automate forensic processes and improve investigative accuracy \cite{bendiab2024forensic}. A recent study highlights how LLM invocation logging enhances forensic readiness, ensuring transparency and auditability in cloud investigations \cite{galhotra2024logging}. Despite these advancements, challenges persist in ensuring the integrity and reliability of cloud logs, which are critical for forensic investigations \cite{mishra2012cloudforensics}. Additionally, as cloud environments grow more complex, forensic methodologies must evolve to address emerging threats and vulnerabilities \cite{zawoad2013cloudforensics}.

New research has explored LLM-based automation for cloud forensic investigations. LLMCloudHunter \cite{schwartz2024llmcloudhunter} introduces a framework that utilizes Large Language Models (LLMs) to automatically generate detection rules from unstructured cloud-based cyber threat intelligence (CTI) sources. Similarly, LogPrécis \cite{boffa2023logprecis} applies LLMs to Unix shell attack logs, extracting attacker tactics and reducing large datasets into manageable forensic fingerprints for enhanced investigation. These studies demonstrate the growing role of LLMs in automating forensic analysis, significantly improving the speed and efficiency of threat detection.

\subsection{Beyond Forensics: AI in Cybersecurity Operations}

Beyond forensic analysis, AI-enhanced Risk-Based Access Management (RBAM) systems have been developed to dynamically adjust permissions by analyzing access logs, leading to reduced false positives and unauthorized access incidents \cite{agorbia2024leveraging}. Additionally, SecGenAI, a framework for securing cloud-based GenAI applications, has been proposed to enhance privacy compliance and mitigate adversarial threats in cloud security \cite{secgenai_cloud_genai_security}. However, while these advancements underscore the potential of GenAI in cybersecurity, they also raise concerns regarding its dual-use nature—where the same technology can be exploited for malicious purposes \cite{dual_use_genai_cybersecurity}.

As AI and LLM-based forensic solutions evolve, threat modeling remains a key strategy in mitigating cybersecurity risks. Threat modeling is a structured approach to identifying, assessing, and mitigating security threats to a system, application, or network. It involves defining assets, recognizing potential threats, analyzing attack vectors, assessing risks, and implementing security controls \cite{verma2024operationalizingthreatmodelredteaming}. This process is particularly crucial in LLM-driven cloud forensics, as AI systems introduce unique vulnerabilities that must be accounted for in forensic investigations.

\section{Cloud Investigation Automation Framework (CIAF)}
\label{CIAF}

Our framework follows the cloud forensic process that involves six steps, identification of an event, identification of evidence, collection of evidence, analysis of evidence, interpretation of results, and presentation of results. The process starts with the identification of an event in which potential incidents or suspicious activities are detected through monitoring systems and alerts. This is followed by the identification of evidence, which involves determining what data and artifacts are relevant to the investigation, such as logs, files, or network traffic. In the evidence collection phase, investigators gather these data while ensuring their integrity and authenticity by creating forensic images or securing logs. During evidence analysis, forensic tools and techniques are used to examine the data in detail, identify patterns, extract relevant information, and reconstruct the sequence of events. Interpretation of results involves evaluating the analyzed data to draw meaningful conclusions, such as determining the attack vector, identifying the perpetrator, or assessing the extent of the impact. Finally, the presentation of results compiles the findings into a comprehensive report, often including visualizations and clear explanations, which is then shared with stakeholders, potentially serving as evidence in legal proceedings or guiding remediation efforts. \cite{purnaye2021}. Cyber experts usually perform the steps manually. However, our framework seeks to automate the cloud forensics process for cyberattacks.
\begin{figure*}[t]
    \centering
    \includegraphics[width=0.8\textwidth]{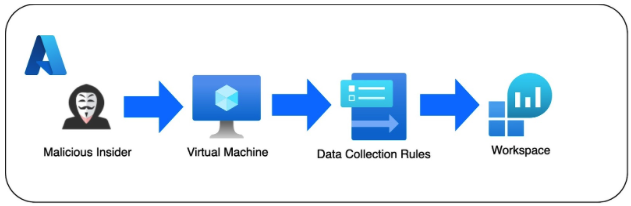}
    \caption{Experiment setup}
    \label{fig:setup}
\end{figure*}

For the evaluation of each respective attack classification, we used confusion matrices, which offer a detailed view of classification results by showing how a model’s predictions align with actual class labels. The matrix displays the counts of true positives (TP), true negatives (TN), false positives (FP), and false negatives (FN), which form the basis for key evaluation metrics such as accuracy, precision, recall, and F1 score. These abbreviations (TP, TN, FP, FN) are used for simplicity in the related formulas.
Precision, recall, and F1 scores are essential metrics for assessing a classification model’s performance, particularly in distinguishing positive and negative classes. Precision indicates the likelihood that a positive prediction by the model is correct. Recall measures the proportion of actual positive instances that the model successfully identified. The F1 score, as the harmonic mean of precision and recall, offers a balanced metric that considers both aspects. Accuracy, on the other hand, is more suitable for balanced datasets and reflects how often the model's predictions match the actual outcomes \cite{inproceedings}.

\begin{align}
\text{Precision} & = \frac{TP}{TP + FP}  \tag{1} \\%[10pt]
\text{Recall} & = \frac{TP}{TP + FN}  \tag{2} \\%[10pt]
F1 \, \text{score} & = 2 \times \frac{\text{Precision} \times \text{Recall}}{\text{Precision} + \text{Recall}}  \tag{3} \\%[10pt]
\text{Accuracy} & = \frac{TP + TN}{TP + FP + TN + FN}  \tag{4}
\end{align}

\vspace{12pt} % Adds 10pt vertical space

Figure 1 shows how our framework performs in the cloud forensic process. The identification of an event and identification of evidence locate and describe the source of the data to be analyzed. In the collection of evidence step, we select the required data, including the event and the evidence, and delimit it by time range.  The framework starts when the user selects the attack as input for our system. In the analysis of evidence step, our framework takes the user input to query the information needed to detect the specified attack. The information includes preprocessed methods and the prompts for the specified attack. The preprocessed method filters the important features and provides steps to shape the data ready to work as input in the LLM. In the interpretation of the results step, the system and user prompt take their structure from the ontology, and then the user prompt takes preprocessed data to perform a query to the LLM. Finally, in the presentation of the results, the LLM provides the classification as final evidence of whether the attack occurred or not.

\begin{figure*}[t]
    \centering
    \includegraphics[width=0.8\textwidth]{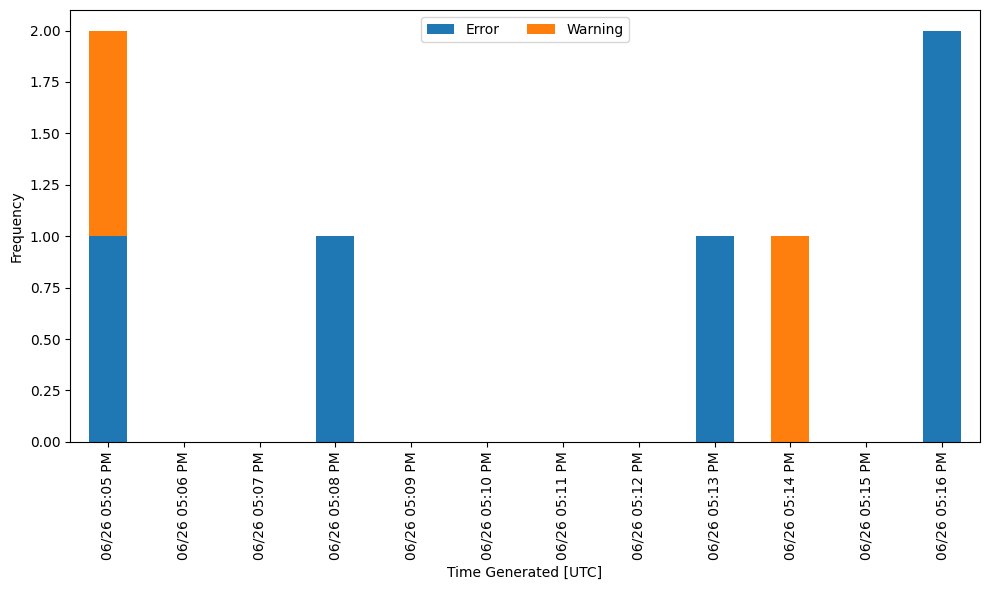} 
    \caption{Event distribution for warnings and errors when a ransomware is successfully launched in Azure VM.}
    \label{fig:Event} % Label for referencing in text
\end{figure*}

In conclusion, our framework streamlines and automates the cloud forensic process for cyberattacks by integrating the traditional six-step methodology - event identification, evidence identification, evidence collection, evidence analysis, result interpretation, and result presentation - into a cohesive and efficient system. By allowing users to input specific attack scenarios, the framework automates data collection, preprocessing, and querying, leveraging large language models (LLMs) to analyze and interpret evidence. This approach not only enhances the efficiency and accuracy of forensic investigations but also reduces the manual effort required by cybersecurity experts.

\begin{table*}[!htbp]
\centering
\caption{Monitored Azure Performance Counters and System Metrics}
\begin{tabular}{|l|p{12cm}|}
\hline
\textbf{Performance Counter} & \textbf{Description} \\
\hline
Thread Count & The number of threads currently running in the system. \\
\% Free Space & The percentage of free disk space available on the system. \\
Working Set - Private & The amount of RAM used exclusively by a process, not shared with others. \\
Processor Frequency & The current clock speed of the processor in MHz. \\
Packets Received Errors & The number of network packets received with errors. \\
Packets Outbound Errors & The number of network packets sent with errors. \\
Working Set & The total amount of physical memory used by a process. \\
Free Megabytes & The amount of free physical memory in MB. \\
Pool Nonpaged Bytes & The size of the non-paged memory pool that remains in RAM. \\
Pool Paged Bytes & The size of the paged memory pool, which can be written to disk. \\
Available Bytes & The amount of available memory that can be allocated immediately. \\
\% Committed Bytes In Use & The percentage of committed virtual memory in use. \\
Processor Queue Length & The number of threads waiting for processor time. \\
Processes & The number of active processes running on the system. \\
Committed Bytes & The total virtual memory committed for use. \\
Handle Count & The number of system object handles used by a process. \\
Cache Bytes & The amount of memory used by the system cache. \\
System Up Time & The total time the system has been running since the last reboot. \\
Avg. Disk Write Queue Length & The average number of write requests waiting for disk access. \\
Avg. Disk Queue Length & The average number of both read and write disk requests queued. \\
Disk Writes/sec & The number of write operations performed per second on a disk. \\
\% User Time & The percentage of processor time spent on user-mode operations. \\
Disk Transfers/sec & The total number of read and write operations per second. \\
Disk Reads/sec & The number of read operations performed per second on a disk. \\
Avg. Disk Read Queue Length & The average number of read requests waiting for disk access. \\
Context Switches/sec & The number of times the processor switches between threads per second. \\
\% Privileged Time & The percentage of processor time spent on system (kernel) operations. \\
Avg. Disk sec/Read & The average time in seconds to read from the disk. \\
\% Processor Time & The percentage of total processor time used. \\
Bytes Sent/sec & The number of bytes sent over the network per second. \\
Bytes Received/sec & The number of bytes received over the network per second. \\
Packets/sec & The total number of network packets sent and received per second. \\
Bytes Total/sec & The total number of bytes sent and received over the network per second. \\
Avg. Disk sec/Transfer & The average time in seconds to complete a disk transfer operation. \\
Packets Sent/sec & The number of network packets sent per second. \\
Disk Bytes/sec & The total number of bytes read and written to the disk per second. \\
Disk Read Bytes/sec & The number of bytes read from the disk per second. \\
\% Idle Time & The percentage of time the processor or disk is idle. \\
\% Disk Write Time & The percentage of time the disk is occupied. \\
\hline
\end{tabular}
\label{tab:azure_system_metrics}
\end{table*}

%%%%%%%%%%%%%%%%%%%%%%%%%%%%%%

\section{Case Study: Automated Ransomware Investigation in Azure}
\label{AzureCaseStudy}

To demonstrate our system, we created a proof of concept as an experiment in Microsoft Azure for detecting Ransomware attacks. The setup is shown in Figure \ref{fig:setup}. 
First, we created a Windows Virtual Machine to serve as the target system. A Log Analytics workspace is also set up, which will store performance logs and security-related data. The VM was then connected to this workspace through Azure Monitor using data collection rules to ensure that all relevant system activities were being recorded as logs for forensic investigation. The data collection rule and the workspace allowed the gathering of all available event information and performance counter features. Event information provides the following features: TenantId, SourceSystem, TimeGenerated [UTC], Source, EventLog, Computer, EventLevel, EventLevelName, ParameterXml, EventData, EventID, RenderedDescription, AzureDeploymentID, Role, EventCategory, UserName, Message, MG, ManagementGroupName, Type, ResourceId, only TimeGenerated [UTC] and EventLevelName is relevant for this specific analysis.
In the case of performance features, Table \ref{tab:azure_system_metrics} shows the list of all collected features.

\begin{figure*}[t]
%\vskip 0.2in
\begin{center}
\centerline{\includegraphics[width=\textwidth]{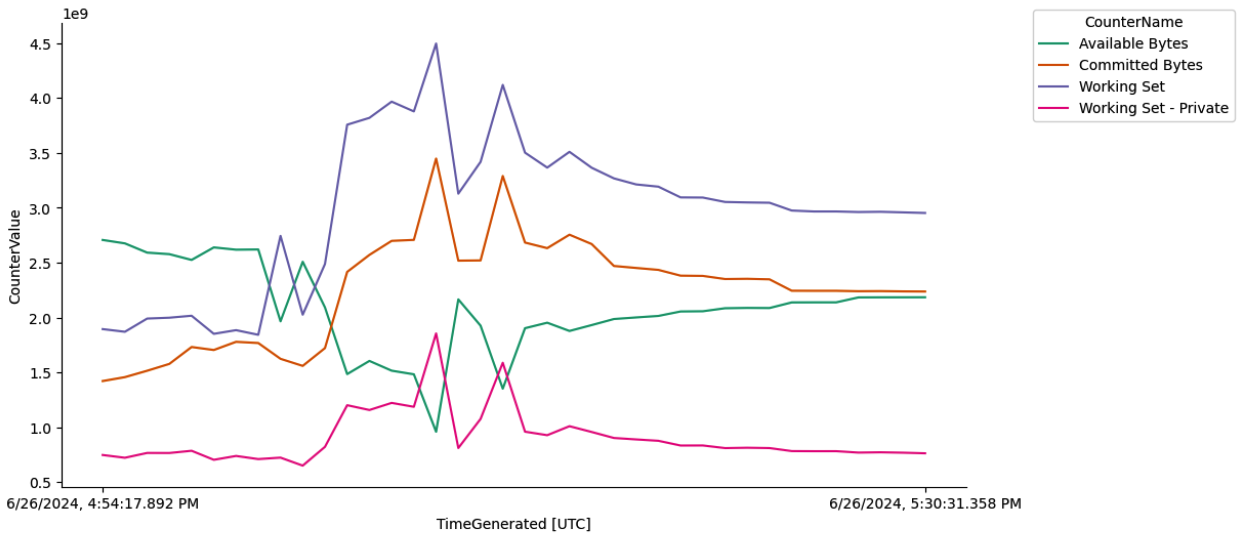}}
\caption{Timeline of Working Set, Working Set - Private, Committed Bytes, and Available Bytes feature behaviors.}
\label{timeline}
\end{center}
\vskip -0.2in
\end{figure*}

To simulate a malicious insider attack, a ransomware attack script was executed to encrypt files and demonstrate the impact of a real-world malware infection. These attacks are designed to trigger security logs, which helps in forensic analysis. While the attack was running, we manually saved time by labeling the data further and comparing it with the LLM results. Once the attack is executed, data logs are examined using Azure Monitor Queries. The Perf and Event queries help validate that log collection is active and allow analysts to filter critical security events. In our analysis, we used the Perf query to obtain the data to be analyzed by an LLM to classify the behavior on the VM.

On the other hand, Microsoft Azure Event query was be used to analyze event data collected from a VM under a ransomware attack. By leveraging Azure Monitor and Log Analytics, you can query event logs for signs of suspicious activities such as unusual login attempts, file system changes (e.g., file encryption), or abnormal network traffic. Using Kusto Query Language (KQL), you can search for specific patterns, like error or warning events related to file access or encryption attempts.

Following the cloud forensics steps, we took the data when the attack occurred, events were detected, we analyzed 30 minutes of the distribution of warning and error, which can be observed in Figure 4. It contains 1692 instances, per feature and minute. 
Based on the standard deviation, we selected important features, this is because those with higher standard deviation are the most affected when the VM behavior changes. The metrics Working Set, Working Set - Private, Committed Bytes, and Available Bytes were isolated to detect anomalies caused by ransomware execution. Then, we preprocessed the selected data by mapping numeric values in those columns to a Likert scale (Very Low, Low, Normal, High, Very High) based on the mean and standard deviation of the column. For each value in the column, it is classified into one of these categories depending on how far it deviates from the mean in terms of standard deviations. After that, we performed data cleaning and transformation by using pivot operation as a column name for the process name. The timeline of events in Figure FF shows when the attack was executed, and how the performance counter features were affected.

The framework uses an ontology to create a system of knowledge where a user can query for known attacks, and the ontology provides the information required for the LLM to perform analysis. The ontology is expandable for unknown attacks after a cybersecurity expert updates the information. When we ask the system to analyze the data to detect the ransomware attack, the ontology retrieves the features and the prompts related to the specific attack scenario. 
Because our system uses LLMs for classification, we need to convert feature values to text. Then, once the important features were selected, we used the 3 Sigma Rule to obtain Likert scale labels (\textit{extremely low, very low, low, normal, high, very high, extremely high}), which is a statistical principle that describes the distribution of data in a normal distribution. The 3 Sigma Rule states that approximately 68\% of data points fall within one standard deviation (\(\pm 1\sigma\)) of the mean, 95\% within two standard deviations (\(\pm 2\sigma\)), and 99.7\% within three standard deviations (\(\pm 3\sigma\)).

\begin{figure}[t]
    \centering
    \includegraphics[width=0.8\columnwidth]{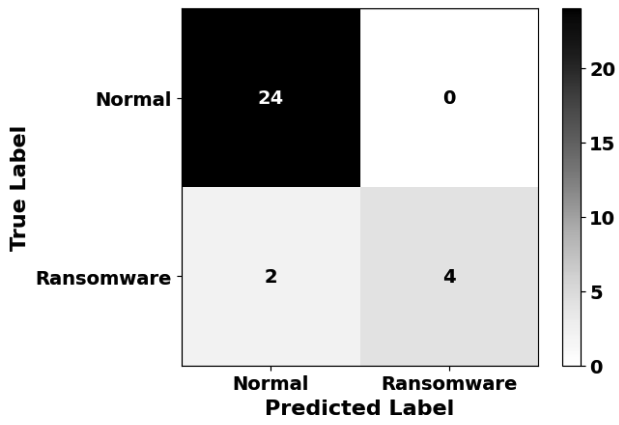} 
    \caption{Confusion matrix of ransomware attack against VM Azure virtual a machine.}
    \label{fig:confusionMatrix} % Label for referencing in text
\end{figure}

The transformed Likert values will be the input to be analyzed by the LLM. We used system and user prompt to obtain our prediction, the system prompt was 'You are a cyber forensics assistant capable of detecting ransomware by applying data analyst techniques, to detect ransomware AvailableBytes should be at least Low and Working Set, WorkingSetPrivate, CommittedBytes should be at least High', and the user prompt ‘based on {data}, classify as normal or ransomware, just provide the classification’. Where data is each of the rows in our dataset. The classification results in Figure \ref{fig:confusionMatrix} of how well the model distinguishes between Normal system behavior and Ransomware activity. One of the key metrics, precision, indicates how many of the predicted positives were correct.

Finally, we evaluated the model’s performance by comparing the actual ransomware labels with the LLM’s predictions. A classification report is generated, detailing precision, recall, and F1-score metrics in Table \ref{tablePerformance}. The model achieved a precision of 0.92 for Normal cases, meaning that 92\% of the instances classified as "Normal" were truly normal. For ransomware, the precision was 1.00, signifying that every instance predicted as ransomware was indeed ransomware, with no false positives.

%%%%%%%%%%%%%%%%%%%%%%%%%%%%%%

\section{Discussion and Future Directions}
\label{discussion}

The experimental evaluation of our AI-driven forensic framework demonstrates significant improvements in investigation efficiency and accuracy. For the case of a successful ransomware in the cloud, the framework drastically reduce the amount of data to be used as input for the LLM. Basically, through the ontology it keeps only relevant features effected by the ransomware attack, then the llm can respond wheter or not the attack occurred . These findings align with prior work emphasizing automation in cloud forensics to cope with the scale and complexity of cloud environments \cite{martini2012cloud}. The different directions for this research are described in the following subsections.

\subsection{real-time implementation}
Our framework can be also implemented in real-time analysis. It has the capability of further reducing the task of forensics because it will work on those successful threats. To implement the framework for real-time analysis, the system would integrate with live monitoring tools to continuously detect suspicious events and automatically collect relevant evidence, such as logs and network traffic. The framework would preprocess this data and use a large language model (LLM) to analyze and interpret it in real-time. The results would be immediately presented on a dashboard, providing cybersecurity teams with actionable insights and enabling rapid response to potential threats. This automation enhances threat detection and minimizes the impact of cyberattacks by streamlining the forensic process.

\subsection{Generalizing Beyond Ransomware}

The proposed framework for automating the cloud forensic process was tested in the context of ransomware attacks. One promising direction is expanding the framework to support a broader range of cyberattacks beyond ransomware, such as insider threats, and advanced persistent threats. By incorporating attack-specific preprocessing methods and refining the ontology, the framework becomes a more versatile tool for comprehensive cloud security investigations.

Another critical area for future work is enhancing the framework integration with diverse cloud environments and platforms. This could involve developing connectors and APIs for seamless interaction with major cloud service providers, enabling more efficient data collection and evidence analysis. Additionally, exploring containerized deployments or serverless architectures could improve the scalability and performance in dynamic cloud ecosystems.

The LLMs in the analysis and interpretation phases present opportunities for future research. Fine-tuning LLMs with domain-specific datasets related to emerging ransomware tactics could improve detection accuracy. 
By continuously evolving with the threat landscape, the framework could provide a more proactive defense against sophisticated cloud-based cyberattacks.

\subsection{Challenges and Open Research Questions}
We identify several key directions for future work to advance AI-driven forensic automation and security challenges:
\begin{itemize}
    \item \textbf{Integration of LLMs for Contextual Analysis:} LLMs have recently shown promise in a variety of security applications, from malware analysis to intrusion detection \cite{xu2024LLM}. Incorporating LLM-based components could enable the framework to interpret unstructured data (such as incident reports, attacker chat logs, or natural-language event descriptions) and perform high-level reasoning. For example, an LLM fine-tuned on cybersecurity incident narratives might assist in piecing together a timeline or hypothesizing attacker objectives from disparate clues.
\end{itemize}
\begin{itemize}
    \item \textbf{Data Quality:} Access to high-quality and diverse cloud forensic logs is often restricted due to privacy, security, and proprietary concerns from cloud service providers. Many providers limit the visibility into their infrastructure, which can hinder comprehensive forensic investigations. Additionally, forensic logs may vary widely in quality, presenting challenges such as missing data, inconsistent timestamps, or obfuscated information. Collaborations with cloud service providers to access synthetic or anonymized datasets could prove beneficial in addressing these challenges \cite{pichan2025}. 
\end{itemize}
\begin{itemize}
    \item \textbf{LLMs accuracy:} LLMs may generate false positives or negatives, especially when dealing with ambiguous log entries or novel attack patterns. The dynamic nature of cloud environments demands that models adapt quickly to emerging threats and evolving log formats. One potential solution is fine-tuning LLMs using domain-specific datasets, including labeled forensic logs and incident reports. Combining LLM outputs with rule-based systems or traditional machine learning models through ensemble techniques could also enhance accuracy \cite{hu2021loralowrankadaptationlarge}. 
\end{itemize}
\begin{itemize}
    \item \textbf{Automation vs. Human Oversight:} Balancing automation with human oversight is crucial in cloud forensic investigations. While automation via LLMs increases efficiency, complete reliance on AI could lead to potential blind spots, particularly in high-stakes environments where forensic conclusions have legal or regulatory implications. In such contexts, human expertise is essential to validate findings and maintain forensic soundness. Adopting a hybrid approach, where LLMs perform preliminary analysis and human analysts review high-priority or ambiguous cases, could optimize the investigation process \cite{sambana2023}. 
\end{itemize}

\begin{table}[t]
\label{tablePerformance}
\centering
\caption{Classification Report}
\begin{tabular}{lcccc}
\hline
Class & Precision & Recall & F1-Score & Support \\
\hline
Normal & 0.92 & 1.00 & 0.96 & 24 \\
Ransomware & 1.00 & 0.67 & 0.80 & 6 \\
Accuracy &  & & 0.93 & 30 \\
Macro Avg & 0.96 & 0.83 & 0.88 & 30 \\
Weighted Avg & 0.94 & 0.93 & 0.93 & 30 \\
\hline
\end{tabular}
\label{tab:classification_report}
\end{table}

\begin{itemize}
    \item \textbf{Regulatory and Compliance Concerns:} Cloud forensic investigations operate within a complex regulatory framework, requiring adherence to data protection regulations such as GDPR, CCPA, and industry-specific standards like HIPAA or PCI DSS. Additionally, cross-jurisdictional investigations must navigate differing local laws and data sovereignty issues. Non-compliance with these legal standards could result in forensic evidence being inadmissible in court or lead to regulatory penalties. To mitigate these risks, the framework should integrate privacy-preserving techniques such as anonymization and pseudonymization \cite{Seth2024Compliance}. 
\end{itemize}
%%%%%%%%%%%%%%%%%%%%%%%%%%%%%%

\section{Conclusion}
\label{conc}

In this paper, we introduced the Cloud Investigation Automation Framework (CIAF) and demonstrated its effectiveness in enhancing ransomware detection through experimental evaluation on Microsoft Azure logs. The framework achieved high precision, recall, and F1 scores, underscoring its potential to significantly improve the accuracy and efficiency of cloud forensic investigations.

Beyond its immediate application in ransomware scenarios, CIAF offers a scalable and adaptable approach to forensic automation across diverse cloud environments. Its ontology-driven standardization of forensic inputs establishes a robust foundation for AI-driven forensic reasoning, reducing dependence on manual analysis while enhancing reproducibility and interpretability.

This research opens new avenues for advancing AI-driven forensic automation. Key questions remain, such as how to integrate causal reasoning mechanisms to enhance attack attribution, how well multi-cloud forensic frameworks can generalize across varied cloud architectures, and what theoretical limits exist for structured semantic validation in forensic investigations. Addressing these challenges will involve further exploration of ontology-guided AI models, enhancing adversarial resilience in forensic systems, and developing real-time forensic decision-making capabilities. By pursuing these directions, future research can push the boundaries of forensic automation, contributing to more robust and reliable cloud security practices.

%%%%%%%%%%%%%%%%%%%%%%%%%%%%
%\section*{References}

\sloppy

\bibliographystyle{unsrtnat}   % or plainnat with [numbers] option above
\bibliography{references, output}

\end{document}